\documentclass[10pt]{article}

\usepackage{amsfonts}
\usepackage{amsmath}
\usepackage{amsthm}
\usepackage{amssymb}
\newtheorem{theorem}{Theorem}[]
\newtheorem{lemma}[theorem]{Lemma}

\theoremstyle{definition}
\newtheorem{definition}[theorem]{Definition}

\newtheorem{remark}[theorem]{Remark}

\newlength{\fuyasu}
\theoremstyle{definition}

\renewcommand\P{{\mathcal P}}
\newcommand\Q{{\mathcal Q}}
\newcommand\K{{\mathcal K}}

\newcommand\geh{\mathfrak{g}}
\newcommand{\gehn}{\ensuremath{\geh_n}}

\newcommand\ol{\overline}

\newcommand\xb{\overline{x}}
\newcommand\yb{\overline{y}}
\newcommand\zb{\overline{z}}
\newcommand\Bb{\overline{B}}
\newcommand\Rb{\overline{R}}
\newcommand\Z{\mathbb{Z}}
\newcommand\Zn{\Z_{\ge0}}

\renewcommand{\thesection}{{\Roman{section}}}

\title{{\textbf{Analysis of a particle antiparticle description \\
of a soliton cellular automaton}}}

\author{Taichiro Takagi\\
\normalsize
\em Department of Applied Physics, National Defense Academy,\\
\normalsize
\em Kanagawa 239-8686, Japan}

\date{}
\begin{document}
\maketitle

\begin{abstract}
We present a derivation of a formula 
that gives dynamics of an
integrable cellular automaton associated with crystal bases.
This automaton is related to type $D$ affine Lie algebra and
contains usual box-ball systems as a special case.
The dynamics is described by means of
such objects as carriers, particles, and antiparticles.
We derive it from an analysis of a recently obtained
formula of the combinatorial $R$ (an intertwiner between
tensor products of crystals) that was found in a study of
geometric crystals.
\end{abstract}

\section{{INTRODUCTION}}\label{sec:1}
The crystal basis theory
\cite{K1,K2} has {played} an important role
in studies of solvable lattice models and integrable systems
{since} more than a decade.
In this context
Ref.~\cite{KKM} by Kang, Kashiwara and Misra has
provided useful families of
crystal bases associated with affine Lie algebras 
$A^{(1)}_n, B^{(1)}_n, C^{(1)}_n, D^{(1)}_n,A^{(2)}_{2n-1}, A^{(2)}_{2n}$ 
and $D^{(2)}_{n+1}$.
We call the first algebra
{\em type $A$} and the fourth one {\em type $D$}.
In Ref.~\cite{HHIKTT} we have obtained
a formula of the {\em combinatorial $R$} (an intertwiner of the crystals)
associated with the type $A$ algebra.
{}From the viewpoint of integrable systems
an intriguing fact {in Ref.~\cite{HHIKTT}} is that
this formula has been derived from a discrete soliton equation
(the nonautonomous discrete KP equation) by a procedure known as the
ultradiscretization.
Since then there has been 
progress which goes beyond the type $A$ case.
In a study of geometric crystals
associated with the type $D$ algebra
we have obtained an explicit formula of a {\em tropical $R$},
an intertwiner of geometric crystals \cite{KOTY1}.
{The tropical $R$ is a birational map between
totally positive rational functions, while the combinatorial $R$
is a bijective map between finite sets.}
Further analysis of this tropical $R$ and the combinatorial $R$ 
derived from it
should be an important task in
studies of integrable systems, since
they are connected with discrete
and ultradiscrete
soliton equations
of type $D$ Lie algebra symmetry \cite{KOTY2}.

The purpose of this paper is 
to investigate a {\em piecewise linear formula}
of the above mentioned combinatorial $R$ for the type $D$ crystals 
in Ref.~\cite{KOTY1}.
{Our main result is the derivation of the} 
limit of the formula that
leads to the
{\em particle antiparticle description}
of an integrable cellular automaton
(Theorem \ref{th:dec5_2}).
This description was recently obtained \cite{KTT} 
by using a 
{\em factorization of the combinatorial $R$
into Weyl group operators},
a property that had been found and proved
in Ref.~\cite{HKT2}.
We emphasize that
the result 
proves a non-trivial fact that 
the factorization of
the combinatorial $R$ can be conducted
in two different ways, 
via the Weyl operator description and 
via the piecewise linear formula.
This point is new 
even in the type $A$ case (Theorem \ref{th:dec5_1}).

We briefly explain 
{the background to our problem.}
There were studies on one dimensional cellular automata
known as the box-ball systems \cite{T,TM,TS,TTM,TTMS}.
It was found that dynamics in
these automata was controlled by
the combinatorial $R$ of the type $A$ crystals \cite{FOY,HHIKTT}.
Based on the families of crystals in Ref.~\cite{KKM}
integrable cellular automata associated with crystals
of the other types were also constructed \cite{HKOTY,HKT1}.
A question about
such generalized automata arose as to whether
we could give a description of their dynamics as
box-ball like systems.
{To answer this question the particle antiparticle description
was found \cite{HKT3,KTT}.}

{In Ref.~\cite{KTT} it was found that the automata associated with
crystal bases of any types of affine Lie algebra in Ref.~\cite{KKM} can be
embedded into the type $D$ case.
Thus one can obtain the particle antiparticle description
of these automata from that of the type $D$ case.
This is the reason why we
devote ourselves into this particular case.}


The plan of this paper is as follows.
In Sec.~\ref{sec:2} the automaton associated with the type $A$ crystals
is reviewed.
The piecewise linear formula of the combinatorial $R$ is
presented.
The particle description of the automaton is 
proved in terms of the piecewise linear formula.
In Sec.~\ref{sec:3} we discuss the piecewise linear formula of 
the combinatorial $R$ of the type $D$ crystals.
Its reduction to the type $A$ case is also shown.
The automaton associated with the type $D$ crystals
is explained in Sec.~\ref{sec:4}.
The formula of a factorized dynamics of an inhomogeneous automaton 
(Theorem \ref{th:dec5_2}) is reviewed.
This formula is proved in Sec.~\ref{sec:5} using
the piecewise linear formula of the combinatorial $R$.
Proofs of several lemmas are given in the Appendix.

\section{{\mathversion{bold} 
$\mathsf{A^{(1)}_{n-1}}$ CASE}}\label{sec:2}
\subsection{{\mathversion{bold} Combinatorial $\mathsf{R}$}}
\label{subsec:nov27_1}
We begin with type $A$ case.
Instead of $A^{(1)}_{n}$
we adopt
$A^{(1)}_{n-1}$ crystals because it enables us
to compare the results with those in the
$D^{(1)}_{n}$ crystals.
For the {notation} 
we use overlines to 
distinguish the symbols from those in the
type $D$ case,
writing $\Bb$ for a crystal, $\Rb$ for the
combinatorial $R$ and so on.

As a set the $A^{(1)}_{n-1}$ crystal 
$\Bb_l$ ($l$ is any positive integer) is given by
\begin{equation}
\Bb_l = \left\{ (x_1,\cdots,x_n) \in
\Zn^{n} \bigg| \sum_{i=1}^n x_i= l 
\right\}.
\end{equation}
The other properties of this crystal are available in Ref.~\cite{KKM}.
In this paper we use no other property of $\Bb_l$.
We will write simply $\Bb$ or $\Bb'$ for $\Bb_l$ with 
arbitrary $l$.
\begin{definition}\label{def:nov24_1}
Given {a pair of variables}
$x=(x_1,\ldots,x_n) \in \Bb, y=(y_1,\ldots,y_n) \in \Bb'$,
{let $\Rb:(x,y) \mapsto (x',y')$ be the piecewise linear map
defined by}
$x'=(x'_1,\ldots,x'_n), y'=(y'_1,\ldots,y'_n)$ where
\begin{align*}
x_i' &= y_i + P_{i+1} - P_i,\\
y_i' &= x_i + P_i - P_{i+1}.
\end{align*}
Here $P_i$ is given by
\begin{equation}\label{eq:nov24_2}
P_i = 
\max_{1 \leq j \leq n}
\left( \sum_{k=1}^{j-1} (y_{k+i-1} - x_{k+i-1}) + 
y_{j+i-1} \right).
\end{equation}
The indices herein involved are interpreted in modulo $n$.
\end{definition}
Except for the {notation} this formula
is the same  one as defined
in Proposition 4.1 of Ref.~\cite{HHIKTT}.
The normalization of (\ref{eq:nov24_2}) is so chosen as the 
$P_1$ to take the same expression as the formula
in Theorem 5.1 of Ref.~\cite{KKM}.
The property of $\Rb$ to intertwine {the} actions of Kashiwara operators
in the 
crystal basis theory 
{was essentially proved in}
Sec.~1 of Ref.~\cite{KOTY1}.
It ensures that the $(x',y')$ falls into
$\Bb' \times \Bb$.

{We note that 
there exist other ways to present the combinatorial $R$
(e.g.~Refs.~\cite{HKOT} and \cite{NY}) but which are not used in this paper.}
\subsection{{Automaton}}
Now we consider the automaton.
In Ref.~\cite{HHIKTT} the space of automaton extends infinitely towards
both ends
\begin{displaymath}
\cdots \times \Bb_{l_{i-1}} \times \Bb_{l_i} 
\times \Bb_{l_{i+1}} \times \cdots.
\end{displaymath}
For our purposes it is sufficient to consider a finite size system like
\begin{equation}\label{eq:nov25_1}
\Bb_{l_{1}} \times \cdots \times \Bb_{l_N}.
\end{equation}
Let {$L(\gg \sum_{i=1}^N l_i)$} be an integer.
We call a particular letter for the ground state of the automaton
a {\em vacuum}.
We can use any letter in $\{1,\ldots,n\}$ 
as a vacuum of the automaton \cite{KTT}.
Throughout this paper we adopt $n$ as the letter for the vacuum.
As in Ref.~\cite{HKT2} we define 
\begin{displaymath}
\Bb_L[n] = \left\{ (x_1,\ldots,x_n)\in \Bb_L \bigg|
x_n \gg x_a \, \mbox{for any}\, a \ne n \right\}.
\end{displaymath}
We write $\Bb \times \Bb' \simeq \Bb' \times \Bb$
for correspondence by $\overline{R}$.
Take any $x \in \Bb_L[n]$.
Applying $\Rb$ successively we have
\begin{align}
\Bb_L[n] \times \left(
\Bb_{l_{1}} \times \cdots \times \Bb_{l_N} \right) & \simeq
\left( \Bb_{l_{1}} \times \cdots \times \Bb_{l_N} \right)
\times \Bb_L[n],\nonumber\\
x \times Y & \mapsto X' \times y',\label{eq:nov9_1}
\end{align}
that gives the following.
\begin{definition}\label{def:dec22_1}
The time evolution operator $\ol{T}$ of the automaton is given by
\begin{displaymath}
\ol{T}: Y \mapsto X'.
\end{displaymath}
\end{definition}
It means that 
we regard $Y$ and $X'$ in (\ref{eq:nov9_1}) as two automaton states
before and after the time evolution.
We note that the operator $\ol{T}$ actually depends on $x$.
\subsection{{Particle description}}\label{subsec:nov30_1}
There is
an interpretation of the automaton that we call a
particle description.
It is the description of the box-ball systems in 
Refs.~\cite{T,TM,TS,TTM}.
Suppose we have balls with index {$a \, (1 \leq a \leq n-1)$}
that we call {\em $a$-balls}.
For $x=(x_1,\ldots,x_n) \in \Bb_{l_i}$ we associate
a box of capacity $l_i$ that has
$x_a$ $a$-balls $(1 \leq a \leq n-1)$
in it.
Then an element of $\Bb_{l_{1}} \times \cdots \times \Bb_{l_N}$
is regarded as a one dimensional array of boxes of
capacities $l_1, \ldots , l_N$ with these balls.
For any $a \,(1 \leq a \leq n-1)$
we consider a carrier of $a$-balls that we call an {\em $a$-carrier}.
We assume that the $a$-carrier has a sufficiently large
capacity, {so that} it can carry arbitrary number of
$a$-balls at a time.

First we suppose $l_i=1$ for all $i$.
We call the associated automaton {\em basic} \cite{KTT}.
In this case 
the $x$ represents a box with an $a$-ball if $x_a=1$
for $a \ne n$.
It represents an empty box if $x_n=1$.
{For any $a$ we}
write \fbox{$\mathstrut a$} for $x$ with $x_a=1$. 
The carrier goes along the array of boxes.
Then there are four actions
in the loading-unloading process 
by the $a$-carrier:
\begin{enumerate}
\item If the carrier has at least one ball and meets an empty box,
we unload a ball from the carrier and put it into the box.
\item If the carrier meets a box with an $a$-ball, 
we pick up the ball and load it into the carrier.
\item If the carrier meets a box with a $b$-ball $(b \ne a)$, we do
nothing.
\item If the carrier has no ball and meets an empty box, we do
nothing.
\end{enumerate}
These actions are depicted by the left four pictures
in Fig.~1.
For any $a \, (1 \leq a \leq n-1)$ let $\K_a$ be
a {\em particle motion operator} that 
acts on the space of automaton (array of boxes)
and does the actions in
the loading-unloading process explained above.
We assume that the $\K_a$ depends on $x$ in (\ref{eq:nov9_1})
in such a way that
the $a$-carrier has $x_a$ balls in it at the beginning
where $x_a$ is the $a$-th element of the $x$.
Then for the basic automaton we have \cite{FOY,HHIKTT}
\begin{displaymath}
\ol{T} = \K_1 \K_2 \ldots \K_{n-1}.
\end{displaymath}

Now we consider a not necessarily basic case which we call
{\em inhomogeneous} \cite{KTT}.
We denote by $\P$ {the} operator that reduces the automaton into
a basic one, and by $\Q$ {the} operator that makes a
rearrangement of balls \cite{KTT}.
To explain them we first let $N=1$ in (\ref{eq:nov9_1})
\begin{displaymath}
\Bb_L[n] \times \Bb_{l}   \simeq \Bb_{l} \times \Bb_L[n].
\end{displaymath}
Then let
\begin{displaymath}
\P: \Bb_l \rightarrow \underbrace{\Bb_1 \times \cdots \times \Bb_1}_{l}
\end{displaymath}
be {the} operator which sends $y=(y_1,\ldots,y_n) \in \Bb_{l}$ into
\begin{displaymath}
\underbrace{
\fbox{$\mathstrut n$} \times \cdots \times \fbox{$\mathstrut n$} }_{y_n}
\times \cdots \times
\underbrace{\fbox{$\mathstrut 1$} \times \cdots \times 
\fbox{$\mathstrut 1$}}_{y_1} .
\end{displaymath}
Its inverse $\P^{-1}$ can be defined only on 
{such arrays in which the letters are arranged in} decreasing order.
Let
\begin{displaymath}
\Q: \underbrace{\Bb_1 \times \cdots \times \Bb_1}_{l}
\rightarrow \underbrace{\Bb_1 \times \cdots \times \Bb_1}_{l}
\end{displaymath}
be {the} operator which packs \fbox{$\mathstrut n$}'s into the left end.
Next we consider {the case $N > 1$}.
We insert walls between the
{$\Bb_{l_i}$'s} in (\ref{eq:nov25_1}) to mark their positions.
Then by $\P$ we denote {the} operator that applies the above 
$\P$ on each $\Bb_{l_i}$, and 
by $\Q$ or $\P^{-1}$ {those} 
that applies the above $\Q$ or $\P^{-1}$
on each $\Bb_1 \times \cdots \times \Bb_1$ between the walls.
Now we have \cite{F,KTT} the following.
\begin{theorem}\label{th:dec5_1}
The time evolution operator of the inhomogeneous automaton is given by
\begin{equation}\label{eq:nov10_2}
\ol{T} = \P^{-1} \Q \K_1 \K_2 \ldots \K_{n-1} \P.
\end{equation}
\end{theorem}
This theorem means 
that the time evolution of the inhomogeneous automaton can be
reduced into that of the basic one only by inserting 
a simple rearrangement.
\subsection{{Proof of the particle description}}
The reduction of an inhomogeneous automaton to a basic one
(Theorem \ref{th:dec5_1}) was first {presented} by Fukuda \cite{F}. 
We gave another proof of this theorem
(and its generalization to type $D$ case)
in Ref.~\cite{KTT}. 
Here we show still another proof of this theorem
that uses the piecewise linear
formula in Definition \ref{def:nov24_1}.
This is the proof that was inferred in Sec.~II D of Ref.~\cite{HHIKTT}
but was not explicitly given there.
Let $p_i = \lim_{x_n \rightarrow \infty} P_i$.
Then by (\ref{eq:nov24_2}) we have
\begin{equation}\label{eq:nov24_3}
p_i= 
\max_{1 \leq j \leq n+1-i}
\left( \sum_{k=1}^{j-1} (y_{k+i-1} - x_{k+i-1}) + 
y_{j+i-1} \right).
\end{equation}
{}From Definition \ref{def:nov24_1} and (\ref{eq:nov24_3})
we obtain $p_n = y_n$ and
\begin{align}
p_i &= \max \{ y_i,y_i-x_i+p_{i+1} \},\label{eq:nov25_3}\\
x'_i &= \min \{ p_{i+1},x_i \}.\label{eq:nov25_4}
\end{align}
Note that the relation (\ref{eq:nov25_3}) 
is a descending recursion formula for {$p_i$'s} on $i$.
Let $x \in \Bb_L[n]$ and $y \in \Bb_l$ 
{be a pair of variables}.
Let $\K_a$ be the particle motion operator 
introduced in Sec.~II.~C.
Then it is easy to see that the {$p_i$'s (respectively $x'_i$'s)}
for $1 \leq i \leq n-1$
obtained by (\ref{eq:nov25_3}) (respectively by (\ref{eq:nov25_4}) ) 
{denote} the number of {\em empty boxes}
(respectively the number of {\em boxes with balls with 
index $i$})
in the automaton state $\K_i \ldots \K_{n-1} \P y$.
We also see that the $a$-carrier finally has $y'_a = x_a+y_a-x'_a$ 
balls in it.
This proves the factorization of {$\overline{T}$} in (\ref{eq:nov10_2}) 
for {the case $N=1$} in (\ref{eq:nov9_1}).
The assertion of the theorem for {the case $N > 1$}
follows immediately
by repeated use of this case, {where we adopt} the
final states of the carriers
for $\Bb_{l_i}$ in (\ref{eq:nov25_1}) as their initial states
for $\Bb_{l_{i+1}}$.
\section{{\mathversion{bold} 
$\mathsf{D^{(1)}_n}$ COMBINATORIAL $\mathsf{R}$}}
\label{sec:3}
\subsection{{Piecewise linear formula}}
\label{subsec:nov29_2}
As a set the $D^{(1)}_n$ crystal 
$B_l$ ($l$ is any positive integer) is given by
\begin{equation}
B_l = \left\{ (x_1,\cdots,x_n,\xb_n,\cdots,\xb_1) \in
\Zn^{2n} \bigg| x_n \xb_n=0, \sum_{i=1}^n (x_i + \xb_i) = l 
\right\}.
\end{equation}
The other properties of this crystal are available in 
a preprint version of 
Ref.~\cite{KKM} (Kyoto Univ., RIMS-887, 1992) 
or in, e.g., Ref.~\cite{KMOTU}.
In this paper we use no other property of $B_l$.
We will write simply $B$ or $B'$ for $B_l$ with 
arbitrary $l$.
\begin{definition}\label{def:nov4_1}
{Let} $x=(x_1,\ldots,\xb_1) \in B, y=(y_1,\ldots,\yb_1) \in B'$
{be a pair of variables}.
{The} involutive automorphisms $*,\sigma_1,\sigma_n$
on $x,y$ {are defined by}
\begin{align*}
* :& \, x_i \leftrightarrow \yb_i,\xb_i \leftrightarrow y_i \quad 
(1 \leq i \leq n),\\
\sigma_1 :& \,x_1 \leftrightarrow \xb_1, y_1 \leftrightarrow \yb_1, \\
\sigma_n :& \,x_n \leftrightarrow \xb_n, y_n \leftrightarrow \yb_n.
\end{align*}
For any function $F=F(x,y)$ we denote by $F^a$ the function
obtained from $F$ by applying {$a = ( *, \sigma_1, \sigma_n)$} to it.
For $x \in B_l$ we write $\ell(x)$ for $l$.
\end{definition}
\begin{definition}\label{def:oct30_1}
Given {a pair of variables}
$x=(x_1,\ldots,\xb_1) \in B, y=(y_1,\ldots,\yb_1) \in B'$,
{let $R:(x,y) \mapsto (x',y')$ be the piecewise linear map
defined by}
$x'=(x'_1,\ldots,\xb'_1), y'=(y'_1,\ldots,\yb'_1)$ where
\begin{equation}\label{eq:oct21}
\begin{array}{llllllll}
x_1' &= & y_1 &+ & V_0^{\sigma_1} &- & V_1,&  \\
x_i' &= & y_i &+ & V_{i-1}&- &V_i + W_i -W_{i-1} & (2 \leq i \leq n-1),\\
x_n' &= & y_n &+ & V_{n-1} &-& V_{n}^{\sigma_n},&  \\
\xb_i' &=&  \yb_i &+& V_{i-1} &-& V_i & (1 \leq i \leq n),\\
%
y_i' &= & x_i &+& V_{i-1}^* &-& V_i^* & (1 \leq i \leq n),\\
\yb_1' &= & \xb_1 &+& V_0^{\sigma_1} &-& V_1^*,&   \\
\yb_i' &= & \xb_i &+& V_{i-1}^* &-& V_i^* +W_i-W_{i-1} & (2 \leq i \leq 
n-1),\\
\yb_n' &= & \xb_n &+& V_{n-1}^* &-& V_{n}^{\sigma_n}.&
\end{array}
\end{equation}
Here $V_i$ and $W_i$ are given by 
\begin{align}
\label{eq:piecewise}
&V_i =  \max_{1 \leq j \leq n-1} \{ \alpha_{i,j},\alpha'_{i,j}  \},
\\
&W_i=\max \left( V_i+V_{i-1}^*-y_i, V_{i-1}+V_i^*-\xb_i \right)
+ \min (x_i, \yb_i ),\; (1 \leq i \leq n-2),  \label{eq:oct24_2}\\
&W_{n-1}=V_{n}+V_{n}^{\sigma_n}. \label{eq:oct24_3}
\end{align}
The functions $\alpha_{i,j}= \alpha_{i,j}(x,y)$ and 
$\alpha'_{i,j}= \alpha'_{i,j}(x,y)$ in (\ref{eq:piecewise}) are given by
\begin{align}
\alpha_{i,j}(x,y) &= \max(\delta_{j,n-1} \beta_i,\yb_j - x_j )
+
\begin{cases}
\ell (x) + \sum_{k=j+1}^i (\yb_k - \xb_k ) & \mbox{for} \, j \leq i,\\
\ell (y) + \sum_{k=i+1}^j (\xb_k - \yb_k ) & \mbox{for} \, j > i,
\end{cases}
\\
\alpha'_{i,j}(x,y) &= \max(\delta_{j,n-1} \beta'_i,x_j - \yb_j )
+\ell (x) + \sum_{k=1}^i (\yb_k - \xb_k )+
            \sum_{k=1}^j (y_k - x_k ),
\end{align}
where
\begin{displaymath}
\beta_i  = 
\begin{cases}
x_n - \yb_n & \mbox{for}\,i \ne n-1,n,\\
0 & \mbox{for}\,i = n-1,\\
\xb_n - y_n & \mbox{for}\,i = n,
\end{cases} \quad
\beta'_i  = 
\begin{cases}
\yb_n - x_n & \mbox{for}\,i \ne n-1,n,\\
\max(y_n - 2 \xb_n, \yb_n - 2 x_n) & \mbox{for}\,i = n-1,\\
y_n - \xb_n & \mbox{for}\,i = n.
\end{cases}
\end{displaymath}
\end{definition}
The map $R$ is the combinatorial $R$ {for} the $D^{(1)}_n$ crystals.
The property of $R$ to intertwine {the} 
actions of Kashiwara operators in the 
crystal basis theory was {proved} in Theorem 4.28 of Ref.~\cite{KOTY1}.
It ensures that the $(x',y')$ falls into $B' \times B$.
\begin{remark}\label{rem:dec16_2}
We have changed {the notation} from Ref.~\cite{KOTY1} since
our {present} formalism uses both $x_n$ and $\xb_n$.
The changes are listed in Table \ref{tab:oct25}.
\end{remark}
\begin{table}[h]
\begin{center}
	\caption{The correspondence of the notation between in this
paper and in Ref.~\cite{KOTY1}
for the piecewise linear
formula of the $D^{(1)}_n$ combinatorial $R$. 
The $z$ in the first column denotes $x,y,x'$ or $y'$.
It is assumed that $i \ne n-1,n$
in the last two columns.}
	\begin{tabular}{|c|c||c|c||c|c|}
		\hline
		\mbox{This paper} & \mbox{Ref.~\cite{KOTY1}}& 
		\mbox{This paper} & \mbox{Ref.~\cite{KOTY1}}& 
		\mbox{This paper} & \mbox{Ref.~\cite{KOTY1}}\\
		\hline
		$z_n$ & $\max(z_n,0)$ &
		$V_n$ & $V_{n-1}$ & 
		$\alpha_{i,j(\ne n-1)}$ & 
		$\max( \theta_{i,j}, \eta_{i,j} )$ \\
		$\zb_n$ & $\max(-z_n,0)$ &
		$V_n^{\sigma_n}$ &  $V_{n-1}^*$ &
		$\alpha'_{i,j(\ne n-1)}$ & 
		$\max( \theta'_{i,j},\eta'_{i,j})$ \\
		$z_{n-1}$ & $z_{n-1}+\min(z_n,0)$ &
		$V_{n-1},V_{n-1}^*$ & --- &
		$\alpha_{i,n-1}$ & 
		$\max( \eta_{i,n-1},\eta_{i,n})$\\
		$\zb_{n-1}$ & $\zb_{n-1}+\min(z_n,0)$ &
		$*$ & $* \circ \sigma_n$ &
		$\alpha'_{i,n-1}$ & 
		$\max( \eta'_{i,n-1},\eta'_{i,n})$\\
		\hline
	\end{tabular}
	\label{tab:oct25}
\end{center}
\end{table}
According to the correspondence in Table \ref{tab:oct25}
{one of the formulas}
in Eq.~(4.66) of Ref.~\cite{KOTY1} is {now} translated
into
\begin{align*}
x'_n 
&= y_n + \max( V_n- \yb_n,V_n^{\sigma_n}-y_n) - V_n^{\sigma_n},\\
\xb'_n &= \yb_n + \max( V_n- \yb_n,V_n^{\sigma_n}-y_n)-V_n.
\end{align*}
In order to make them coincide with the relations
in (\ref{eq:oct21}) we should define
the $V_{n-1}$ as
$\max( V_n- \yb_n,V_n^{\sigma_n}-y_n)$.
Actually the above definition of $V_{n-1}$ is equivalent to this.
In other words we have the following.
\begin{lemma}
The following relation holds:
\begin{equation} \label{eq:oct23_7}
V_{n-1} = \max( V_n - \yb_n,V_n^{\sigma_n} - y_n).
\end{equation}
\end{lemma}
\begin{proof}
We have
\begin{align*}
\alpha_{n-1,j} &= \max( \alpha_{n,j} - \yb_n,
\alpha_{n,j}^{\sigma_n} - y_n),\\
\alpha'_{n-1,j} &= \max( \alpha'_{n,j} - \yb_n,
(\alpha'_{n,j})^{\sigma_n} - y_n),
\end{align*}
for $1 \leq j \leq n-1$.
In order to check these relations we can use
$\max(-x_n,-\xb_n) = \max(-y_n,-\yb_n) = 0$.
The claim of the lemma follows immediately from these relations.
\end{proof}
%
\begin{remark}
{The transformation} properties of the piecewise linear functions
$V_i$ and $W_i$ under the automorphisms $\sigma_1,*$, and $\sigma_n$
will be used afterwards, {so we list}
them in Table \ref{tab:one}.
{It} was quoted from Ref.~\cite{KOTY1}
and adjusted by the correspondence in Table \ref{tab:oct25}.
\end{remark}
\begin{table}[h]
\begin{center}
\caption{The transformation of the piecewise linear functions
$V_i$ and $W_i$ in Definition \ref{def:oct30_1} 
by the automorphisms $\sigma_1,*$, and $\sigma_n$ in
Definition \ref{def:nov4_1}.}
	\begin{tabular}[t]{|c|c|c|c|c|}
		\hline
		& $V_0$ & $V_i \, (1 \leq i \leq n-1)$ & $V_n$ & 
		$W_i \, (1 \leq i \leq n-1)$ \\
		\hline
	$\sigma_1$& $V_0^{\sigma_1}$ & $V_i$ & $V_n$ & $W_i$ \\
	$\ast$ & $V_0$ & $V_i^*$ & $V_n$ & 
	$W_i$ \\
	$\sigma_n$& $V_0$ & $V_i$ & $V_n^{\sigma_n}$ & $W_i$ \\
		\hline
	\end{tabular}
	\label{tab:one}
\end{center}
\end{table}
\noindent
Some more relations on the piecewise linear functions
will be used later.
We give them at the beginning of the Appendix.
\subsection{{\mathversion{bold} Reduction to 
the $\mathsf{A^{(1)}_{n-1}}$ case}}
{
We realize that the piecewise linear formula in type $D$ case
has a rather bulky expression in contrast with its
type $A$ counterpart:
See Definitions \ref{def:nov24_1} and \ref{def:oct30_1}.
In order to understand its
structure
it is worth trying to study
some special limits of the formula.
}
Here we consider a reduction to the type $A$ case. 
We observe that
the piecewise linear map $R$ 
in Definition \ref{def:oct30_1}
for the
$D_n^{(1)}$ crystals
{becomes} the intertwiner of the
$A_{n-1}^{(1)}$ crystals 
under {the reduction}.
\begin{theorem}\label{lem:oct25}
Set
\begin{equation}\label{eq:oct29_1}
\xb_i = \yb_i =0 \quad \mbox{ for } \quad 1 \leq i \leq n.
\end{equation} 
Then the map $R:(x,y) \mapsto (x',y')$ in 
Definition \ref{def:oct30_1} reduces to
\begin{align*}
x_i' &= y_i + P_{i+1} - P_i,\\
y_i' &= x_i + P_i - P_{i+1},\\
\xb_i' &= 0,\\
\yb_i' &= 0,
\end{align*}
where $P_i$ {was defined} by (\ref{eq:nov24_2}).
\end{theorem}
This theorem follows from Lemma \ref{lem:oct29_4} below.
\begin{lemma}\label{lem:oct29_4}
Under the specialization (\ref{eq:oct29_1}) the
following relations hold
\begin{align*}
&V_i = \ell(x) + P_1,
\quad V_i^* = \ell(x) + P_{i+1} \quad (0 \leq i \leq n),\\
&V_0^{\sigma_1} = \ell(x) + P_2, \quad
V_n^{\sigma_n} = \ell(x) + P_n,\\
&
W_i = 2 \ell(x) + P_1 + P_{i+1} \quad (1 \leq i \leq n-1).
\end{align*}
\end{lemma}
We shall give a proof of this lemma in Appendix.

{
We note that the reduction from type
$D$ to type $A$ (Theorem \ref{lem:oct25}) itself can also be
obtained from the description of the combinatorial $R$ 
in Ref.~\cite{HKOT} since the insertion algorithms for types
$A$ and $D$ in Ref.~\cite{HKOT} coincide under the condition (\ref{eq:oct29_1}).}

\section{{\mathversion{bold} 
$\mathsf{D^{(1)}_n}$ AUTOMATON}}\label{sec:4}
\subsection{{Definition}}
We now present a brief definition of the $D^{(1)}_n$ automaton
using the crystals and the combinatorial $R$.
For a more complete definition, see Refs.~\cite{HKOTY} and \cite{HKT1}.
We consider a finite size system like
\begin{displaymath}
B_{l_{1}} \times \cdots \times B_{l_N}.
\end{displaymath}
Let {$L(\gg \sum_{i=1}^N l_i)$} be an integer.
We define 
\begin{equation}\label{eq:dec25_1}
B_L[n] = \left\{ (x_1,\ldots,x_n,\xb_n,\ldots,\xb_1)
\in B_L \bigg|
x_n \gg x_a \, \mbox{for any}\, a \ne n \right\}.
\end{equation}
Take any $x \in B_L[n]$.
Applying the combinatorial $R$ successively we have
\begin{align}
B_L[n] \times \left(
B_{l_{1}} \times \cdots \times B_{l_N} \right) &\simeq
\left( B_{l_{1}} \times \cdots \times B_{l_N} \right)
\times B_L[n],\nonumber\\
x \times Y & \mapsto X' \times y',\label{eq:nov26_1}
\end{align}
that gives the following.
\begin{definition}\label{def:dec22_2}
The time evolution operator $T$ of the automaton is given by
\begin{displaymath}
T: Y \mapsto X'.
\end{displaymath}
\end{definition}
Remarks similar to those 
after Definition \ref{def:dec22_1} also apply here.
\subsection{{Particle antiparticle description}}
\label{subsec:nov30_2}
We consider a
particle antiparticle description of this automaton \cite{HKT3,KTT}.
This is a generalization of the particle description in 
Sec.~\ref{sec:3} \ref{subsec:nov30_1}.
Suppose we have balls with index $a$ and 
$\overline{a}, (1 \leq a \leq n-1)$
that we call an $a$-ball and an {\em $\overline{a}$-ball} respectively.
The  $a$-ball and the $\overline{a}$-ball are regarded as
a particle and an antiparticle one another.
We introduce a pair annihilation process {in which} a pair of
particle and antiparticle makes a bound state, 
and a pair creation process where
the bound state breaks up into a pair of
particle and antiparticle of another kind.
{To each} 
$x=(x_1,\ldots,x_n,\xb_n,\ldots,\xb_1) \in B_{l_i}$ we associate
a box of capacity $l_i$ that has
$x_a$ $a$-balls,  $\xb_a$ $\overline{a}$-balls $(1 \leq a \leq n-1)$,
and $\xb_n$ bound states in it.
Then {any} element of $B_{l_{1}} \times \cdots \times B_{l_N}$
{can be} regarded as a one dimensional array of boxes of
capacities $l_1, \ldots , l_N$ with {the} balls and 
{the} bound states.
For any $a \,(1 \leq a \leq n-1)$
we {introduce the notion of} 
an $a$-carrier as in Sec.~\ref{sec:3} \ref{subsec:nov30_1}, 
and {that of}
a carrier {for} $\overline{a}$-balls that we call 
an {\em $\overline{a}$-carrier}.
Assume that their capacities are sufficiently large.

First we consider a basic case, i.e.~we suppose $l_i=1$ for all $i$.
In this case 
the $x$ represents a box with an $a$-ball if $x_a=1$ and
a box with an $\overline{a}$-ball if $\xb_a=1$ for $a \ne n$.
It represents an empty box if $x_n=1$ and
a box with a bound state if $\xb_n=1$.
We write \fbox{$\mathstrut a$} for $x$ with $x_a=1$ and
write \fbox{$\mathstrut \overline{a}$} for $x$ with $\xb_a=1$.
\begin{remark}\label{rem:dec12_3}
In what follows we write $a$ also for a number with an overline
as well as that without an overline.
We interpret $\overline{\overline{a}}=a$ and
$x_{\overline{a}}=\xb_a$.
We call $\xb_a$ the $\overline{a}$-th element of $x$.
\end{remark}
Besides the four actions in the loading-unloading process 
in Sec.~\ref{sec:3} \ref{subsec:nov30_1} we have
three additional actions by the $a$-carrier:
\begin{enumerate}\setcounter{enumi}{4}
\item If the carrier has at least one ball and meets a box
with an $\overline{a}$-ball,
we unload a ball from the carrier and make a
bound state in the box.
\item If the carrier meets a box with a bound state, 
we extract an $a$-ball from the bound state,
load it into the carrier, and leave an $\overline{a}$-ball
in the box.
\item If the carrier has no ball and meets a box with
an $\overline{a}$-ball, we do nothing.
\end{enumerate}
These actions are depicted by the right three pictures
in Figure 1.

For any $a \in \{1,\ldots,n-1\} \cup \{\ol{n-1},\ldots,\ol{1}\}$ 
(see Remark \ref{rem:dec12_3})
let $\K_a$ be a particle motion operator \cite{HKT3,KTT} that 
acts on the space of automaton and does the actions in
the loading-unloading process explained above.
We assume that the $\K_a$ depends on $x$ in (\ref{eq:nov26_1})
in such a way that
the $a$-carrier has $x_a$ balls in it at the beginning,
where $x_a$ is the $a$-th element of $x$.
Then for the basic automaton we have \cite{HKT3}
\begin{displaymath}
T = \K_{\ol{n-1}} \ldots \K_{\ol{2}} \K_{\ol{1}}
   \K_1 \K_2 \ldots \K_{n-1}.
\end{displaymath}

Now we consider an inhomogeneous case.
We define {the} operators $\P$ and $\Q$ 
as in Sec.~\ref{sec:3} \ref{subsec:nov30_1} but 
modify them to be suitable for the type $D$ case \cite{KTT}.
To {define} them we first {set} $N=1$ in (\ref{eq:nov26_1})
\begin{equation}
B_L[n] \times B_{l}   \simeq B_{l} \times B_L[n].
\end{equation}
Then let
\begin{displaymath}
\P: B_l \rightarrow \underbrace{B_1 \times \cdots \times B_1}_{l}
\end{displaymath}
be {the} operator which 
sends $y=(y_1,\ldots,y_n,\yb_n,\ldots,\yb_1) \in B_{l}$ into
\begin{equation}\label{eq:dec16_1}
\underbrace{
\fbox{$\mathstrut \ol{1}$} \times \cdots \times 
\fbox{$\mathstrut \ol{1}$} }_{\yb_1}
\times \cdots \times
\underbrace{
\fbox{$\mathstrut \ol{n}$} \times \cdots \times 
\fbox{$\mathstrut \ol{n}$} }_{\yb_n}
\times
\underbrace{\fbox{$\mathstrut n$} \times \cdots \times 
\fbox{$\mathstrut n$}}_{y_n}
\times \cdots \times
\underbrace{\fbox{$\mathstrut 1$} \times \cdots \times 
\fbox{$\mathstrut 1$}}_{y_1} .
\end{equation}
Its inverse $\P^{-1}$ can be defined only on 
{such arrays in which the letters are arranged}
as in (\ref{eq:dec16_1}).
Let
\begin{displaymath}
\Q: \underbrace{B_1 \times \cdots \times B_1}_{l}
\rightarrow \underbrace{B_1 \times \cdots \times B_1}_{l}
\end{displaymath}
be {the} 
operator which packs \fbox{$\mathstrut n$}'s into the left end and
\fbox{$\mathstrut \ol{n}$}'s into the right end.
{For the case $N > 1$ we generalize the definitions of
these operators in the same way}
as in Sec.~\ref{sec:3} \ref{subsec:nov30_1}.
Now we have the following.
\begin{theorem}\label{th:dec5_2}
The time evolution operator of the inhomogeneous automaton is given by
\begin{equation}\label{eq:nov29_1}
T = \P^{-1} \K_{\ol{n-1}} \ldots \K_{\ol{2}} \K_{\ol{1}}
\Q \K_1 \K_2 \ldots \K_{n-1} \P.
\end{equation}
\end{theorem}
In Ref.~\cite{KTT} a proof of this theorem
was given
by means of the factorization of the combinatorial $R$ 
in Ref.~\cite{HKT2}.
In the remaining part of this paper
we give another proof of this theorem
that uses the piecewise linear
formula of the combinatorial $R$ in Definition \ref{def:oct30_1}.
\section{{PROOF OF THE PARTICLE ANTIPARTICLE
DESCRIPTION}}\label{sec:5}
\subsection{{Limit of the piecewise linear formula}}
\label{subsec:dec1_2}
We study a limit of
the piecewise linear formula of the map $R$ in 
Definition \ref{def:oct30_1}.
Let $F=F(x,y)$ be any function of $(x,y) \in B \times B'$.
{The limit we consider here} is to adopt the $B_L[n]$ in
(\ref{eq:dec25_1}) as {the} $B$.
We introduce the following normalized limits:
\begin{align}\label{eq:dec11_1}
{\textstyle \lim_\star }F &= \lim_{x_n \to \infty, \xb_n \to 0}(F(x,y) - \ell(x)),
\\
\label{eq:dec12_1}
{\textstyle \lim_{\star \star}}F 
&= \lim_{x_n \to \infty, \xb_n \to 0}(F(x,y) - 2 \ell(x)).
\end{align}
First we consider the limit (\ref{eq:dec11_1}) of $V_i$.
For the sake of notational simplicity
we denote $\lim_\star V_i$ by $v_i$.
For $a=\sigma_1,\sigma_n$ or $*$
we denote $\lim_\star V_i^a$ by $v_i^a$.
Note that if $a=\sigma_n$ or $*$ the
$v_i^a$ is not necessarily equal to the function
that is obtained from $v_i$ by applying $a$ to it,
whereas {if} $a=\sigma_1$ it is.
Next we consider the limit (\ref{eq:dec12_1}) of $W_i$.
We shall denote $\lim_{\star \star} W_i$ by $w_i$.

The relations in Lemma \ref{lem:dec3_1} in the Appendix
become recursion relations
in the limit (\ref{eq:dec11_1}).
{We let $(x)_+$ denote $\max(x,0)$.}
\begin{lemma}\label{lem:oct24}
For $F=V_i,V_i^*,V_0^{\sigma_1}$ or $V_n^{\sigma_n}$
the limit $\lim_\star F$ exists.
Moreover the following relations hold
\begin{align}
v_n^{\sigma_n} &= v_{n-1}^* = y_n - \yb_n,\label{eq:oct27_9}
\\
v_{i-1}^* &= y_i - x_i + \max\{ v_i^*,(x_i - \yb_i)_+\},
\quad (1 \leq i \leq n-1) \label{eq:oct23_4}\\
v_i &= \max\{ \yb_i - \xb_i + v_{i-1}, (\yb_i - x_i)_+ \},
\quad (1 \leq i \leq n-1) \label{eq:oct23_5}\\
v_n &= \yb_n - y_n +
\max\{y_n + \yb_{n-1} - \xb_{n-1} + v_{n-2}, 
(y_n + \yb_{n-1} - x_{n-1})_+ \}.  \label{eq:oct23_6}
\end{align}
\end{lemma}
We consider the limit (\ref{eq:dec12_1}) of the defining relations
of $W_i$ ((\ref{eq:oct24_2}) and (\ref{eq:oct24_3})).
\begin{lemma}\label{lem:dec1_1}
The following relation holds
\begin{equation}\label{eq:oct24_4}
w_i = v_i + v_i^* - \min \{
v_i^* - v_{i-1}^* + y_i, v_i - v_{i-1}+\xb_i \} + 
\min \{ x_i,\yb_i \}
\quad (1 \leq i \leq n-1).
\end{equation}
\end{lemma}
These lemmas will be used in proofs of Lemmas \ref{lem:oct27_2}
and \ref{lem:oct27_3}.
We shall give their proofs in the Appendix.
\subsection{{Analysis of the particle antiparticle 
description}}\label{subsec:dec1_3}
We now consider a recursion formula satisfied by {the}
numbers of items in the particle antiparticle 
description.
For this purpose we introduce the following.
\begin{definition}\label{def:oct29_3}
For {any} non-negative integers
$A,B,C,D,E$, define {the} piecewise linear map
\begin{displaymath}
\gamma : (A,B,C,D,E) \mapsto (F,G,H,I,J)
\end{displaymath}
by
\begin{align*}
& F = \min(A,E), \\
& G = B + (A-E)_+,\\
& H = \min \left( C,B+(E-A)_+ \right), \\
& I = D + \left( C-B-(E-A)_+ \right)_+, \\
& J = D + \left( B-C+(E-A)_+ \right)_+.
\end{align*}
\end{definition}
The identities $F+G=A+B$, $H+I=C+D$, and $F+H+J=B+D+E$
can be checked easily and will be used afterwards.
We give an interpretation of the map $\gamma$ 
in the particle antiparticle description that is illustrated 
in Figure 2.
Recall the seven actions in the loading-unloading process
by the $a$-carrier (Figure 1)
that were explained
in Secs.~\ref{sec:2} \ref{subsec:nov30_1} 
and \ref{sec:4} \ref{subsec:nov30_2}.
We write {\em act-$i$} for the action with number $i$
in {the} lists.
Note that the $a$ can represent {an overlined number} in Figure 1.
In Figure 2 the boxes with $b$-balls ($b \ne 
a,\ol{a}$) have been omitted because of act-3.
In the upper picture of Figure 2 we are
applying act-5 (or act-7 if $E=0$), act-6, 
act-1 (or act-4 if $B+(E-A)_+=0$), and act-2
from left to right.
In the lower picture we are
applying act-1 (or act-4 if $E=0$), act-2, 
act-5 (or act-7 if $B+(E-A)_+=0$), and act-6.

In what follows we always assume
\begin{equation}\label{eq:oct27_8}
x_n \gg 0 \quad \mbox{and} \quad \xb_n=0.
\end{equation}
Let $B$ and $B'$ be the $D^{(1)}_n$ crystals.
\begin{definition}\label{def:oct29_4}
{For each pair of variables}
$x=(x_1,\ldots,\xb_1) \in B, y=(y_1,\ldots,\yb_1) \in B'$,
{define the set of variables}
$z^{(i)}, \zb^{(i)} \, (0 \leq i \leq 2n-2)$,
$y_i^\bigcirc, \yb_i^\bigcirc (1 \leq i \leq n-1), \, 
x'_i,\xb'_i,y'_i,\yb'_i(1 \leq i \leq n)$
as follows.
\begin{enumerate}
\item
Set
\begin{equation}
\zb^{(0)} = \yb_n, \quad z^{(0)} = y_n.
\label{eq:oct27_4}
\end{equation}
\item
Define
$z^{(n-i)}, \zb^{(n-i)},y_i^\bigcirc, \yb_i^\bigcirc \,y'_i
\, (1 \leq i \leq n-1)$
as
\begin{align}
&
\begin{array}{rllllll}
\gamma (& \yb_{i},&\zb^{(n-1-i)},&z^{(n-1-i)},&y_{i},&x_{i}&) \\
=(& \zb^{(n-i)},& \yb_{i}^\bigcirc,& y_{i}^\bigcirc, &z^{(n-i)},&
y'_{i}&),
\end{array}
\label{eq:oct27_5}
\end{align}
by descending recursion on $i$.
Here the function $\gamma$ is given by Definition \ref{def:oct29_3}.
\item
Define
$z^{(n-1+i)}, \zb^{(n-1+i)},x'_i,\xb'_i,\yb'_i
\, (1 \leq i \leq n-1)$
as
\begin{align}
&
\begin{array}{rllllll}
\gamma (&z^{(n-2+i)},&\yb_{i}^\bigcirc, &y_{i}^\bigcirc,&
\zb^{(n-2+i)},&\xb_{i}&)\\
=(& \xb_{i}',&z^{(n-1+i)},& \zb^{(n-1+i)},&x_{i}',&\yb_{i}'&),
\end{array}
\label{eq:oct27_6}
\end{align}
by recursion on $i$.
\item
Set
\begin{align}
&x_n'=z^{(2n-2)},\quad \xb_n'=\zb^{(2n-2)},
\quad y_n'=\ell(x)-\sum_{i=1}^{n-1}(y_i'+\yb_i'),\quad \yb_n'=0.
\label{eq:oct27_7}
\end{align}
\end{enumerate}
\end{definition}
\begin{remark}\label{rem:dec12_2}
{Let us consider the case when} $x \in B_L[n]$ and $y \in B_l$.
Then the numbers represented by the variables $z^{(i)}$ etc.~in 
Definition \ref{def:oct29_4} are equal to the numbers of items
in the particle antiparticle description in 
Sec.~\ref{sec:4} \ref{subsec:nov30_2}.
More precisely these items appear 
within the time evolution process by $T$ in 
Theorem \ref{th:dec5_2} for the {case $N=1$}.
See Table \ref{tab:nov30_1}.
In {the} table $t_i$ and $\ol{t}_i$ are defined as follows:
For $1 \leq i \leq n-1$ we let
$t_i$ (respectively $\overline{t}_i$)
{denote} the time just after the $i$-carrier 
(respectively $\overline{i}$-carrier) has passed,
where the automaton state is given by $\K_i 
\cdots \K_{n-1} \P y$
(respectively $\K_{\ol{i}}\cdots\K_{\ol{1}}\Q
\K_1 \cdots \K_{n-1} \P y$).
\end{remark}
\begin{table}[h]
\caption{The correspondence between the variables in 
Definition \ref{def:oct29_4} and the items in
the particle antiparticle description.}
\label{tab:nov30_1}
		\begin{tabular}[t]{|c|c||c|c|}
			\hline
Variables	& Items at time $t_i$&
Variables  & Items at time $\overline{t}_i$\\
			\hline
$z^{(n-i)}$			&empty boxes &
$z^{(n-1+i)}$ & empty boxes \\
$\zb^{(n-i)}$			&boxes with bound states &
$\zb^{(n-1+i)}$ & boxes with bound states \\
$y_{i}^\bigcirc$		&boxes with $i$-balls &
$x'_i$ & boxes with $i$-balls \\
$\yb_{i}^\bigcirc$		&boxes with $\ol{i}$-balls &
$\xb'_i$ & boxes with $\ol{i}$-balls \\
$y'_i$	 & balls in the $i$-carrier &
$\yb'_i$ & balls in the $\ol{i}$-carrier \\
			\hline
		\end{tabular}
\end{table}
\subsection{{Proof}}
We now give the proof of Theorem \ref{th:dec5_2} that we have
promised at the end of Sec.~\ref{sec:4}.
It is obtained from the following.
\begin{theorem}\label{th:oct27_1}
{Let $x=(x_1,\ldots,\xb_1) \in B, y=(y_1,\ldots,\yb_1) \in B'$
be a pair of variables,
and} suppose the condition (\ref{eq:oct27_8}) on $x$.
Let $v_i,w_i,v^*_i,v^{\sigma_1}_0, v^{\sigma_n}_n$ be {the}
functions defined in Sec.~\ref{sec:5} \ref{subsec:dec1_2}, and
let $x'_i,\xb'_i,y'_i,\yb'_i (1 \leq i \leq n)$ be {the} variables
given by Definition \ref{def:oct29_4}.
Then the following relations hold
\begin{equation}\label{eq:oct27_10}
\begin{array}{llllllll}
x_1' &= & y_1 &+ & v_0^{\sigma_1} &- & v_1,&  \\
x_i' &= & y_i &+ & v_{i-1}&- &v_i + w_i -w_{i-1} & (2 \leq i \leq n-1),\\
x_n' &= & y_n &+ & v_{n-1} &-& v_{n}^{\sigma_n},&  \\
\xb_i' &=&  \yb_i &+& v_{i-1} &-& v_i & (1 \leq i \leq n),\\
%
y_i' &= & x_i &+& v_{i-1}^* &-& v_i^* & (1 \leq i \leq n),\\
\yb_1' &= & \xb_1 &+& v_0^{\sigma_1} &-& v_1^*,&   \\
\yb_i' &= & \xb_i &+& v_{i-1}^* &-& v_i^* +w_i-w_{i-1} & (2 \leq i \leq 
n-1),\\
\yb_n' &= & \xb_n &+& v_{n-1}^* &-& v_{n}^{\sigma_n}.&
\end{array}
\end{equation}
\end{theorem}
%
A proof of Theorem \ref{th:oct27_1}
will be given after the following two lemmas.
\begin{lemma}\label{lem:oct27_2}
Let $A=\yb_{i},B=\min(x_{i+1},\yb_{i+1}),
C=v_i^*+\min(x_{i+1},\yb_{i+1}),D=y_{i}$ and $E=x_{i}$
in $\gamma : (A,B,C,D,E) \mapsto (F,G,H,I,J)$.
Then 
\begin{align*}
F &= \min(x_{i},\yb_{i}),\\
G &= \yb_{i} - \min(x_{i},\yb_{i})+\min(x_{i+1},\yb_{i+1}),\\
H &= - \min(x_{i},\yb_{i})+\min(x_{i+1},\yb_{i+1})
+ y_{i}+v_{i}^*-
v_{i-1}^*,\\
I &= v_{i-1}^* + \min(x_{i},\yb_{i}),\\
J &= x_{i} + v_{i-1}^* -v_{i}^*,
\end{align*}
for $1 \leq i \leq n-1$.
\end{lemma}
In what follows we {set $w_0 = 2v_0$}.
Note that we have $v_0^*=v_0$ from Table \ref{tab:one}
and $w_1=v_0+v_0^{\sigma_1}$ from Lemma \ref{lem:dec12_4}
in the Appendix.
\begin{lemma}\label{lem:oct27_3}
Let $A=v_{i-1}+\min(x_i,\yb_i),
B=\yb_i-\min(x_i,\yb_i)+\min(x_{i+1},\yb_{i+1}),
C=-\min(x_i,\yb_i)+\min(x_{i+1},\yb_{i+1})+y_i+v_i^*-v_{i-1}^*,
D=\min(x_i,\yb_i)+v_{i-1}+v_{i-1}^*-w_{i-1}$, and $E=\xb_{i}$
in $\gamma : (A,B,C,D,E) \mapsto (F,G,H,I,J)$.
Then
\begin{align*}
F &= \yb_i+v_{i-1}-v_i,\\
G &= v_i+ \min(x_{i+1},\yb_{i+1}),\\
H &= \min(x_{i+1},\yb_{i+1}) + v_i + v_i^* -w_i,\\
I &= y_i+w_i-w_{i-1}-v_i+v_{i-1},\\
J &= \xb_i+w_i-w_{i-1}-v_i^*+v_{i-1}^*,
\end{align*}
for $1 \leq i \leq n-1$.
\end{lemma}
We shall give proofs of these lemmas in the Appendix.
\begin{proof}[Proof of Theorem \ref{th:oct27_1}]
Suppose $i=n-1$ in Lemma \ref{lem:oct27_2}.
Then we have $B=\yb_n=\zb^{(0)}$ and $C=y_n=z^{(0)}$
because of (\ref{eq:oct27_9}), (\ref{eq:oct27_8}), and (\ref{eq:oct27_4}).
Then by comparing (\ref{eq:oct27_5}) with Lemma \ref{lem:oct27_2}
we see that 
$\zb^{(1)},\yb_{n-1}^{\bigcirc},y_{n-1}^{\bigcirc},z^{(1)}$, and
$y_{n-1}'$ in (\ref{eq:oct27_5})
should be equal to $F,G,H,I$, and $J$.
Thus the expression for $y_{n-1}'$ was obtained.
The expressions for $y'_i \, (1 \leq i \leq n-2)$,
as well as those for 
$\zb^{(n-i)},\yb_i^{\bigcirc},y_i^{\bigcirc},z^{(n-i)}$
will be obtained by descending recursion on $i$, where one uses
$F$ and $I$ as $B$ and $C$ in the next step.

Since $B$ and $C$ in Lemma \ref{lem:oct27_3} are equal to
$G$ and $H$ in Lemma \ref{lem:oct27_2},
we have $B=\yb_i^{\bigcirc}$ and $C=y_i^{\bigcirc}$
in Lemma \ref{lem:oct27_3}.
We also see that when $i=1$ we have $A=z^{(n-1)}$ and
$D=\min(x_1,\yb_1)=\zb^{(n-1)}$ in Lemma \ref{lem:oct27_3}, from the
result obtained in the preceding paragraph.
Then by comparing (\ref{eq:oct27_6}) with Lemma \ref{lem:oct27_3}
we see that $\xb_1',z^{(n)},\zb^{(n)},x_1'$,
and $\yb_1'$ in (\ref{eq:oct27_6})
are equal to 
$F,G,H,I$, and $J$ in Lemma \ref{lem:oct27_3} if $i=1$.
Thus the expressions for $x_1',\xb_1',y_1'$ were obtained.
The expressions for $x_i',\xb_i',y_i' \, (2 \leq i \leq n-2)$
will be obtained by recursion on $i$, where one uses
$G$ and $H$ as $A$ and $D$ in the next step.

Then from (\ref{eq:oct27_7}) we can obtain
the expressions for $x_n'$ and $\xb_n'$ in (\ref{eq:oct27_10})
since we have (\ref{eq:oct27_8}) and (\ref{eq:oct27_9}).
It is clear that the relation
$\yb_n'=0=\xb_n + v_{n-1}^* - v_{n}^{\sigma_n}$ 
holds.
Then the expression for $y_n'$ in (\ref{eq:oct27_10})
{is obtained} from the condition $\ell(y')=\ell(x)$.
The proof is completed.
\end{proof}
{Finally we give the proof of Theorem \ref{th:dec5_2}.}
\begin{proof}[Proof of Theorem \ref{th:dec5_2}]
If we impose the condition (\ref{eq:oct27_8}) on the defining relations
(\ref{eq:oct21}) in Definition \ref{def:oct30_1}
then {their right hand sides 
become those of (\ref{eq:oct27_10})
because of the existence of the limiting functions defined in
Sec.~V \ref{subsec:dec1_2}.}
Then Theorem \ref{th:oct27_1} tells that 
the numbers represented by
$x'_i,\xb'_i,y'_i,\yb'_i$ in Definition \ref{def:oct30_1}
are equal to those by the same symbols in
Definition \ref{def:oct29_4} under this condition.
Then according to Remark \ref{rem:dec12_2}
we see that
the time evolution $T$ by Definition \ref{def:dec22_2}
is identical to the $T$ in Theorem \ref{th:dec5_2} 
for {the case $N=1$}.
The assertion for {the case $N > 1$} follows immediately
by repeated use of this case, where we {adopt} the
final states of the carriers
for $B_{l_i}$ in (\ref{eq:nov26_1}) as their initial states
for $B_{l_{i+1}}$.
\end{proof}

\vspace{0.4cm}
\noindent
{\textbf{ACKNOWLEDGMENTS}} \hspace{0.1cm}
The author thanks Atsuo Kuniba and Akira Takenouchi for
a collaboration in the previous work.

\setcounter{section}{0}
\renewcommand{\thesection}{{APPENDIX} :}
\setcounter{equation}{0}
\renewcommand{\theequation}{A \arabic{equation}}
\section{{PROOFS OF THE LEMMAS}}\label{app:a}
Before giving the proofs
we present some relations between 
piecewise linear functions.
They are used in the main
text and in Appendix.
These relations have been
{essentially} obtained in Ref.~\cite{KOTY1}.

{Ultradiscretization \cite{TM,TTM,TTMS}
is a procedure to derive an equation of piecewise linear
functions from an equation of
totally positive (i.e.~having no minus sign) rational functions.
It is realized as a transformation that replaces $+,\times$ and $/$ by
$\max (\min),+$ and $-$, respectively.}
As {the} ultradiscretization of Lemma 4.12 of Ref.~\cite{KOTY1} we have
the following.
\begin{lemma}\label{lem:dec12_4}
The following relation holds
\begin{equation}
W_1 = V_0 + V_0^{\sigma_1}.
\end{equation}
\end{lemma}
This lemma is used {just above} Lemma \ref{lem:oct27_3}.
The next lemma is obtained from {the formulas}
(4.23)$^*$,(4.23), and (4.24) of
Ref.~\cite{KOTY1} by {the} ultradiscretization.
\begin{lemma}\label{lem:dec3_1}
The following relations hold
\begin{align}
&\max\{ V^*_i ,\ell(x), \ell(x) + x_i -\yb_i \} =
\max\{ x_i - y_i + V^*_{i-1},\ell(y),\ell(y)+ x_i -\yb_i \}, 
\label{eq:oct23_1}\\
&\max\{ V_i ,\ell(y), \ell(y) + \yb_i -x_i \} =
\max\{ \yb_i - \xb_i + V_{i-1},\ell(x),\ell(x)+ \yb_i -x_i \}, 
\label{eq:oct23_2}\\
&\max\{ V_n ,\ell(y)+X \}
= \max\{ \yb_{n-1}+\yb_n-\xb_{n-1}-\xb_n + V_{n-2},
\ell(x)+X \},\label{eq:oct23_3}
\end{align}
where $1 \leq i \leq n-2$ and
$X = \yb_n-y_n+
(\yb_{n-1}+y_n-x_{n-1}-\xb_n)_+$.
\end{lemma}
{Here we write $(x)_+$ for $\max(x,0)$.}
This lemma will be used in the proof of Lemma \ref{lem:oct24}.
\subsection*{{Proof of Lemma \ref{lem:oct29_4}}}
\begin{proof}
We derive the expression for $V_i$.
First we suppose $i \ne n-1,n$.
Then under the specialization (\ref{eq:oct29_1}) we have
$V_i =  \max_{1 \leq j \leq n-1} \{ \alpha_{i,j},\alpha'_{i,j}  \}$ with
\begin{align*}
\alpha_{i,j} &= x_n \delta_{j,n-1}+
\begin{cases}
\ell(x) & \mbox{for $j \leq i$},\\
\ell(y) & \mbox{for $j > i$},
\end{cases}
\\
\alpha'_{i,j} &= \ell (x) + \sum_{k=1}^{j-1} (y_k - x_k )+y_j.
\end{align*}
Since $\alpha_{i,j(\leq i)} \leq \alpha'_{i,1}$ and
$\alpha_{i,j(> i)} \leq \alpha_{i,n-1}$, we can drop off
$\alpha_{i,j(\ne n-1)}$ in the max.
Thus we obtain the desired result from (\ref{eq:nov24_2}) with $i=1$.
Now we suppose $i =n-1$ or $n$.
Then we have
\begin{align*}
\alpha_{i,j} &= \ell(x)+
\begin{cases}
0 & \mbox{for $i=n-1$},\\
\max(-y_n,-x_{n-1}) & \mbox{for $i=n$},
\end{cases}
\\
\alpha'_{i,j} &= \ell (x) + \sum_{k=1}^{j-1} (y_k - x_k )+y_j
+ \delta_{j,n-1} \max(y_n-x_{n-1},0) .
\end{align*}
Since $\alpha_{i,j} \leq \alpha'_{i,1}$, we can drop off
$\alpha_{i,j}$ in the max and
obtain the desired result.

We derive the expression for $V_i^*$.
Note that
if $i=0,n$ it has already been proved since $V_i^* = V_i$
for $i=0,n$.
Suppose $i \ne 0,n$.
Then under the specialization (\ref{eq:oct29_1}) we have
$V_i^* =  \max_{1 \leq j \leq n-1} \{ \alpha_{i,j}^*,
(\alpha'_{i,j})^*  \}$ with
\begin{align*}
\alpha_{i,j}^* &= x_j +
\begin{cases}
\ell (y) + \sum_{k=j+1}^i (x_k - y_k ) & \mbox{for} \, j \leq i,\\
\ell (x) + \sum_{k=i+1}^j (y_k - x_k ) & \mbox{for} \, j > i,
\end{cases}
\\
(\alpha'_{i,j})^* &= \delta_{j,n-1} x_n
+\ell (y) + \sum_{k=1}^i (x_k - y_k ).
\end{align*}
Since $(\alpha'_{i,j(\ne n-1)})^* \leq \alpha'_{i,1}$, we can drop off
$(\alpha_{i,j (\ne n-1)})^*$ in the max.
The remaining candidates are
\begin{align*}
\alpha_{i,i+1}^* &= \ell(x) + y_{i+1},\\
\alpha_{i,i+2}^* &= \alpha_{i,i+1}^*  -x_{i+1}+ y_{i+2},\\
& \cdots , \\
\alpha_{i,n-1}^* &= \alpha_{i,n-2}^*  -x_{n-2}+ y_{n-1},\\
(\alpha'_{i,n-1})^* &= \alpha_{i,n-1}^*  -x_{n-1}+ y_{n},\\
\alpha_{i,1}^* &= (\alpha'_{i,n-1})^*  -x_{n}+ y_{1},\\
\alpha_{i,2}^* &= \alpha_{i,1}^* -x_{1}+ y_{2},\\
& \cdots , \\
\alpha_{i,i}^* &= \alpha_{i,i-1}^* -x_{i-1}+ y_{i}.
\end{align*}
Thus we obtain the desired result.

We derive the expression for $V_0^{\sigma_1}$.
Under the specialization (\ref{eq:oct29_1}) we have
$V_0^{\sigma_1} =  \max_{1 \leq j \leq n-1} \{ \alpha_{0,j}^{\sigma_1},
(\alpha'_{0,j})^{\sigma_1}  \}$ with
\begin{align*}
\alpha_{0,j}^{\sigma_1} &= \ell(y)+x_1-
(1-\delta_{j,1})y_1+\delta_{j,n-1}x_n,
\\
(\alpha'_{0,j})^{\sigma_1} &= 
\ell (x) + \sum_{k=2}^j (y_k - x_k ) + (1-\delta_{j,1}) x_j.
\end{align*}
Since $\alpha_{0,j(\ne n-1)}^{\sigma_1} 
\leq \alpha_{0,1}^{\sigma_1}$ and
$(\alpha'_{0,1})^{\sigma_1} 
\leq (\alpha'_{0,2})^{\sigma_1} $, we can drop off
$\alpha_{0,j (\ne 1,n-1)}^{\sigma_1}$ and 
$(\alpha'_{0,1})^{\sigma_1}$ in the max and obtain the desired result.

We derive the expression for $V_n^{\sigma_n}$.
Under the specialization (\ref{eq:oct29_1}) we have
$V_n^{\sigma_n} =  \max_{1 \leq j \leq n-1} \{ \alpha_{n,j}^{\sigma_n},
(\alpha'_{n,j})^{\sigma_n}  \}$ with
\begin{align*}
\alpha_{n,j}^{\sigma_n} &= \ell(x)+y_n-
(1-\delta_{j,n-1})x_n,
\\
(\alpha'_{n,j})^{\sigma_n} &= 
\ell (x) + y_n - x_n + \sum_{k=1}^{j-1} (y_k - x_k ) + y_j.
\end{align*}
Since $\alpha_{n,j(\ne n-1)}^{\sigma_n} 
\leq (\alpha'_{n,1})^{\sigma_n}$, we can drop off
$\alpha_{n,j(\ne n-1)}^{\sigma_n}$ 
in the max and obtain the desired result.

{The expression for $W_{n-1}$ is derived from
(\ref{eq:oct24_3}), and
that for $W_{i(\ne n-1)}$ is from
(\ref{eq:oct24_2}) and the following lemma.}
\end{proof}
\begin{lemma}
{Let $P_i$ be the function that was defined by (\ref{eq:nov24_2}). Then}
\begin{displaymath}
P_{i+1} \geq P_i-y_i.
\end{displaymath}
\begin{proof}
It is easy to see {that}
\begin{displaymath}
P_i-y_i = \max_{1 \leq j \leq n} \{ A_j \}, \quad
P_{i+1} = \max_{2 \leq j \leq n+1} \{ x_i + A_j \}, 
\end{displaymath}
where
\begin{displaymath}
A_j = \sum_{k=1}^{j-1} (y_{i+k} - x_{i+k-1}).
\end{displaymath}
The claim of the lemma holds since we have
$x_i \geq 0$ and $x_i + A_2 = y_{i+1} \geq 0 = A_1$.
\end{proof}

\end{lemma}
\subsection*{{Proof of Lemma \ref{lem:oct24}}}
\begin{proof}
By definition we have 
$V_n^{\sigma_n} = 
\max_{1 \leq j \leq n-1} \{ \alpha_{n,j}^{\sigma_n} ,
(\alpha'_{n,j})^{\sigma_n}   \}$, where
\begin{align*}
(\alpha_{n,j})^{\sigma_n} 
&= \max(\delta_{j,n-1} (x_n - \yb_n) ,\yb_j - x_j )\\
&+ \ell (x) + \sum_{k=j+1}^{n-1} (\yb_k - \xb_k ) 
+ y_n - x_n,\\
(\alpha'_{n,j})^{\sigma_n} 
&= \max(\delta_{j,n-1} (\yb_n - x_n),x_j - \yb_j )\\
&+\ell (x) + \sum_{k=1}^{n-1} (\yb_k - \xb_k )+
            \sum_{k=1}^j (y_k - x_k ) + y_n - x_n.
\end{align*}
In the limit $\lim_\star$ 
the only element that survives in the $\max$ is
$(\alpha_{n,n-1})^{\sigma_n} $, which yields 
$v_n^{\sigma_n} =y_n - \yb_n$.
In the same way the relation
\begin{equation}\label{eq:oct24_1}
v_{n-2}^* = y_{n-1}+y_n-x_{n-1}+
\left( x_{n-1}-\yb_{n-1}+y_n \right)_+,
\end{equation}
can be obtained by a direct calculation.
Then from (\ref{eq:oct23_1}) we see that the $v_i^*$'s exist and
the relation (\ref{eq:oct23_4}) holds for $1 \leq i \leq n-2$ 
by descending induction on $i$.
Since $V_0 = V_0^*$ (Table \ref{tab:one}) we have $v_0 = v_0^*$.
Then from (\ref{eq:oct23_2}) and (\ref{eq:oct23_3}) 
we see that $v_i$'s exist for $i \ne n-1$, and that
the relations
(\ref{eq:oct23_5}) for $1 \leq i \leq n-2$ and
(\ref{eq:oct23_6}) hold, 
by induction on $i$.
Since $v_n$ and $v_n^{\sigma_n}$ exist, we see by (\ref{eq:oct23_7})
that the function $v_{n-1}$ also exists and {equals} to 
$\max \{ v_n  -\yb_n, -\yb_n \}$.
Substituting (\ref{eq:oct23_6}) into 
$v_{n-1}=\max \{ v_n  -\yb_n, -\yb_n \}$ we obtain
(\ref{eq:oct23_5}) for $i=n-1$.
{}From $*$ of (\ref{eq:oct23_7}) we obtain $v_{n-1}^* = y_n - \yb_n$.
Then from (\ref{eq:oct24_1}) and $\max \{ -y_n,-\yb_n \} =0 $
we obtain (\ref{eq:oct23_4}) for $i=n-1$.
{}From $\sigma_1$ of (\ref{eq:oct23_1}) the existence of 
$v_0^{\sigma_1}$ can be verified.
The proof is completed.
\end{proof}
\subsection*{{Proof of Lemma \ref{lem:dec1_1}}}
\begin{proof}
For $1 \leq i \leq n-2$ the relations follow immediately from
(\ref{eq:oct24_2}).
We consider {the case $i=n-1$}.
We obtain $w_{n-1} = v_n + v_n^{\sigma_n}$ from
(\ref{eq:oct24_3}).
First we suppose $x_{n-1} \leq \yb_{n-1}$.
Then from Lemma \ref{lem:oct24} we obtain 
$v_{n-1} = \yb_{n-1} + \max\{ -\xb_{n-1} + v_{n-2},-x_{n-1} \}$.
It yields the relations
$v_n + v_n^{\sigma_n} = v_{n-1} + y_n $ and
\begin{align*}
\mbox{RHS of (\ref{eq:oct24_4})} &= 
v_{n-1} + y_n - \yb_n + x_{n-1} -
\min \{ x_{n-1} - \yb_n, \xb_{n-1} -v_{n-2} - v_{n-1} \} \\
&= v_{n-1} + y_n - \min \{ 0,
\yb_n + (\yb_{n-1} -x_{n-1})+
(\xb_{n-1} -x_{n-1} -v_{n-2})_+ \} \\
&=  v_{n-1} + y_n .
\end{align*}
Thus the assertion of the lemma was proved in this case.
Now we suppose $x_{n-1} > \yb_{n-1}$.
Then
\begin{align*}
\mbox{RHS of (\ref{eq:oct24_4})} &= 
v_{n-1} + y_n - \yb_n + \yb_{n-1} \\
& \qquad -\min \{ x_{n-1} -(x_{n-1}-\yb_{n-1}-y_n)_+
- \yb_n, \xb_{n-1} -v_{n-2} - v_{n-1} \} \\
&= \max \{ (y_n + \yb_{n-1} -x_{n-1} )_+ + v_{n-1},
v_{n-2}+ \yb_{n-1} -\xb_{n-1} +y_n - \yb_n \}.
\end{align*}
{}From Lemma \ref{lem:oct24} we have
$v_{n-1} = ( \yb_{n-1} -\xb_{n-1} + v_{n-2})_+$.
Therefore
\begin{align*}
\mbox{RHS of (\ref{eq:oct24_4})} 
&= \max \left\{ (y_n + \yb_{n-1} -x_{n-1} )_+ , \right.\\
& \qquad \qquad \left. \yb_{n-1} -\xb_{n-1} +y_n + v_{n-2}+
\max\{ -y_n,-\yb_n,\yb_{n-1}-x_{n-1} \} \right\}. 
\end{align*}
The last expression gives $v_n + v_n^{\sigma_n}$ since
the inner max vanishes.
The proof is completed.
\end{proof}
\subsection*{{Proof of Lemma \ref{lem:oct27_2}}}
\begin{proof}
The expressions for $F$ and $G$ are given by definition.
The expression for $I$ is given by $I=C+D-H$, and
that for $J$ is by $J=I+B-C+E-F$.
Thus it suffices to prove $H$.
We have
\begin{align*}
H &= \min(C,B+E-F) \\
&= \min(x_{i+1},\yb_{i+1}) + 
\min\left\{v_i^*,-\min(x_i,\yb_i)+x_i \right\}\\
&= \min(x_{i+1},\yb_{i+1}) +v_i^* - \min(x_i,\yb_i)+x_i-
\max\left\{v_i^*,-\min(x_i,\yb_i)+x_i \right\}.
\end{align*}
Then by (\ref{eq:oct23_4}) we obtain the desired
result.
\end{proof}

\subsection*{{Proof of Lemma \ref{lem:oct27_3}}}
\begin{proof}
The expression for $G$ is given by $G=A+B-F$,
that for $I$ is by $I=C+D-H$, and
that for $J$ is by $J=I+B-C+E-F$.
Thus it suffices to prove $F$ and $H$.
For $F$ we have
\begin{align*}
F &= \min(A,E)\\
&= \yb_i + v_{i-1} + \min\left\{ \min(x_i,\yb_i)-\yb_i,
\xb_i-\yb_i-v_{i-1} \right\}\\
&= \yb_i + v_{i-1} - \max\left\{ -\min(x_i,\yb_i)+\yb_i,
-\xb_i+\yb_i+v_{i-1} \right\}.
\end{align*}
Then by (\ref{eq:oct23_5}) we obtain the desired
result.
For $H$ we have
\begin{align*}
H &= \min(C,B+E-F)\\
&= \min(x_{i+1},\yb_{i+1})-\min(x_{i},\yb_{i})+
\min \{v_i^* - v_{i-1}^* + y_i, v_i - v_{i-1}+\xb_i \}.
\end{align*}
Then by (\ref{eq:oct24_4}) we obtain the desired
result.
\end{proof}

%
\clearpage
\vspace{2cm}
\begin{figure}[htbp]
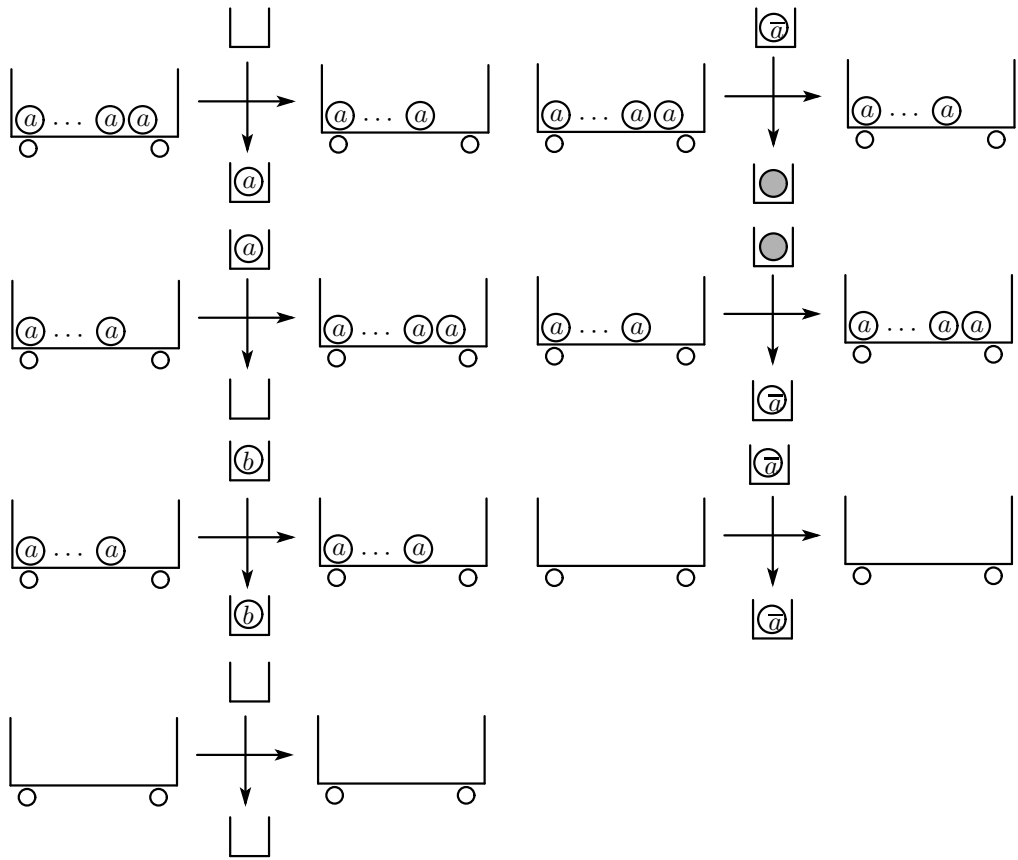

\unitlength 0.1in
%
\caption{The loading and unloading process by
the carrier of balls with index $a$. A gray ball represents {the}
bound state of a particle and an antiparticle.
It is assumed that $b \ne a, \overline{a}$.}
	\label{fig:1}
\end{figure}
\clearpage
\vspace{2cm}
\begin{figure}[htbp]
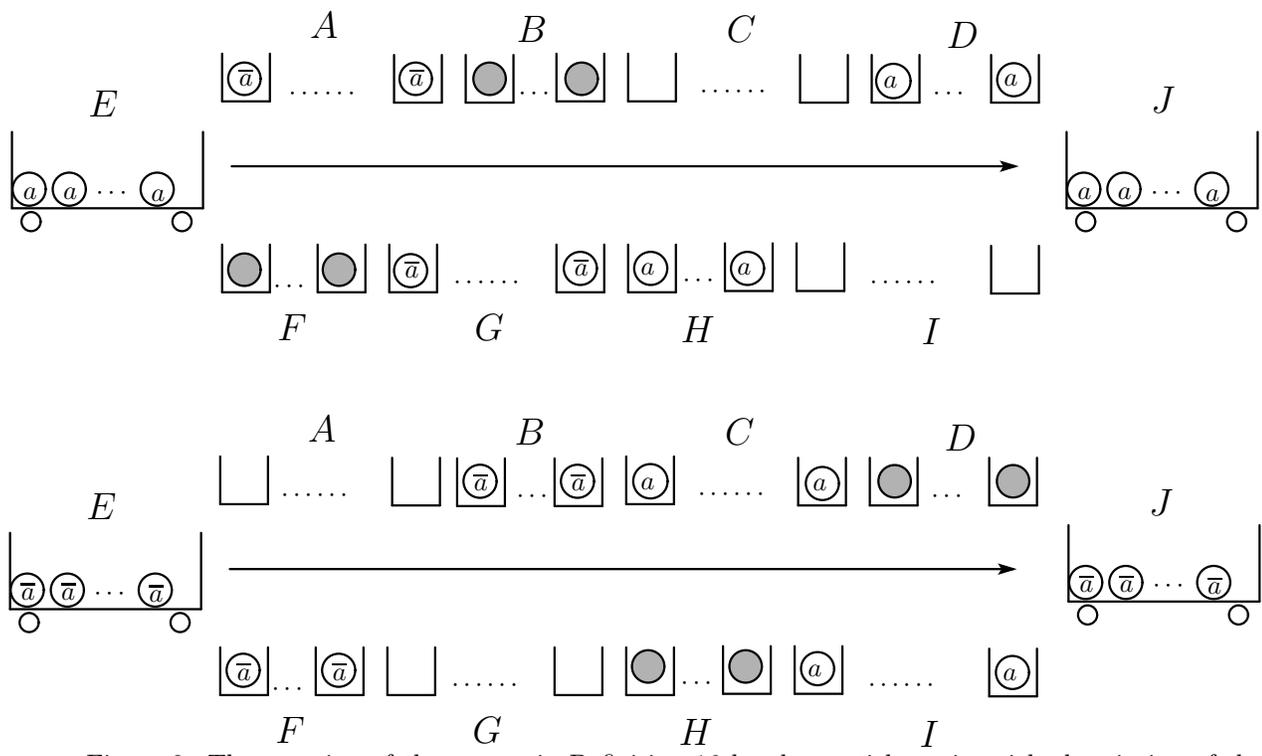

    \hspace*{-4cm}
\unitlength 0.1in
%
%
\caption{The meaning of the map $\gamma$ in Definition \ref{def:oct29_3}
by the particle antiparticle description of the automaton.}
	\label{fig:2}
\end{figure}
\end{document}